\documentclass{optica-article}

\journal{opticajournal}

\articletype{Research Article}
\usepackage{multirow}
\usepackage{bm}
\usepackage{fancyhdr}
\usepackage{booktabs}

\fancypagestyle{empty}{%
  \fancyhf{} 
  \fancyfoot[C]{\footnotesize Approved for public release; distribution is unlimited. Public Affairs release approval \# AFRL-2026-0858 .\\[1ex]
  \normalsize \thepage}
  
}

\begin{document}

\title{ReVAR: A Data-Driven Algorithm for Generating Aero-Optic Phase Screens}

\author{Jeffrey W. Utley,\authormark{1,*} Gregery T. Buzzard,\authormark{1} Charles A. Bouman,\authormark{2} and Matthew R. Kemnetz\authormark{3}}

\address{\authormark{1}Department of Mathematics, Purdue University, West Lafayette, Indiana 47907, USA\\
\authormark{2}Departments of Electrical and Computer Engineering, and Biomedical Engineering, Purdue University, West Lafayette, Indiana 47907, USA\\
\authormark{3}Department of Engineering Physics, Air Force Institute of Technology, Wright-Patterson Air Force Base, Ohio 45433, USA}

\email{\authormark{*}utleyj@purdue.edu}

\begin{abstract*}
    The propagation of light through a turbulent flow field around an aircraft results in optical distortions commonly known as aero-optic effects. The development of methods to mitigate these effects requires large amounts of realistic aero-optic data. However, methods for obtaining this data, including experiment, computational fluid dynamics, and simple phase screen algorithms (e.g., boiling flow), each have significant drawbacks such as high cost, high computation, limited quantity, and/or inaccurate statistics. More recently, data-driven algorithms have been proposed that are computationally efficient and can synthesize aero-optic data to match the statistics of measured data, but these approaches still have drawbacks including limited quality, inaccurate statistics, and the use of complicated algorithms. In this paper, we introduce ReVAR (Re-whitened Vector AutoRegression), a data-driven algorithm for generating synthetic aero-optic data that matches the statistics of measured data. A key contribution in this algorithm is Long-Range AutoRegression, a linear predictive model that combines a standard autoregression with a set of low-pass filters of the data to fit both short-range and long-range temporal statistics. ReVAR uses Long-Range AR together with a spatial re-whitening step to convert measured aero-optic data to temporally and spatially un-correlated white noise. ReVAR can then generate synthetic aero-optic data by reversing this process using white noise input. Using two measured turbulent boundary layer data sets, we demonstrate that ReVAR better matches the measured data's temporal power spectrum and other key metrics than do two conventional phase screen generation methods and an existing single time-lag autoregressive model.
\end{abstract*} 

\section{Introduction}
\label{s: introduction}
Aerodynamic turbulence around an aircraft can interfere with the accurate transmission or reception of light for purposes of range finding or imaging. In particular, atmospheric density variations near the aircraft lead to refractive index variations, causing optical distortions \cite{WangPhysicsComputation, JumperPhysicsMeasurement, Fitzgerald} called aero-optic effects \cite{JumperAero-Optical, SuttonAero, JumperRecentAdvancements}. These effects are quantified as phase aberrations \cite{Visbal, Kalensky, WangPhysicsComputation} and can be measured by optical sensors \cite{Kemnetz, Geary, Holmes} in both wind tunnel experiments \cite{GordeyevFluidic, VukasinovicFlowControl, VukasinovicHybrid} and flight tests \cite{JumperAAOL, JumperAAOL-T}. The resulting data, called aero-optic phase screens, are used to develop predictive control methods that augment adaptive-optic (AO) systems to mitigate aero-optic effects \cite{ShafferPredictive, Sahba, BurnsEstimation}. Specifically, predictive control methods use aero-optic phase screen data to train algorithms such as dynamic mode decomposition \cite{Shaffer, Kutz, ShafferPredictive, Sahba}, neural networks \cite{ShafferNeuralNetwork, BurnsALatency, BurnsARobust}, and autoregressive (AR) models \cite{BurnsEstimation}. Thus, large data sets of aero-optic phase screens are essential to provide adequate training data for these algorithms.

However, obtaining large data sets of aero-optic phase screens through physical experiments is expensive and time-intensive. First, supersonic wind tunnels can measure only very short time series of aero-optic data \cite{GordeyevOpticalCharacterization, GordeyevOpticalMeasurementsTransitional, Berger}. In addition, wind tunnels can introduce non-physical artifacts in the measured data \cite{GordeyevHybrid}, are costly to run, and can simulate only limited aircraft environments \cite{JumperAAOL}. Likewise, generating in-flight data is also expensive, time-intensive, and requires extensive planning and coordination \cite{JumperAAOL}.

As an alternative to measured data, a variety of simulation methods have been investigated to generate synthetic aero-optic phase screens. Computational Fluid Dynamics (CFD) algorithms can simulate aero-optic phase aberrations directly \cite{WangAero-Optics, WangComputation, Porter}, but these algorithms face a strict trade-off between physical relevance and computational efficiency \cite{UtleyBoiling2}. Specifically, high-fidelity CFD is often computationally expensive \cite{GordeyevFluidDynamics, WangPhysicsComputation, GordeyevFluidDynamics2}, while computationally cheaper low-fidelity CFD produces less realistic data \cite{GordeyevFluidDynamics, WangPhysicsComputation}. In addition, because CFD algorithms require physical parameters to be entered by a user, it is difficult to match CFD simulations to measured aero-optic phase screens.

Other phase screen generation methods such as frozen-flow \cite{PoyneerExperimental} and boiling flow \cite{Srinath, UtleyBoiling, UtleyBoiling2} can also generate phase screen data using the Kolmogorov theory of turbulence \cite{Schmidt} and the Taylor frozen-flow hypothesis \cite{Taylor}. These algorithms are computationally inexpensive and can generate data using few parameters. Further, Utley \textit{et al.} \cite{UtleyBoiling2} introduced an algorithm to estimate these parameters from measured aero-optic phase screens. However, because aero-optic effects do not follow the Kolmogorov theory of turbulence \cite{Vogel, Siegenthaler}, the phase screens generated by these methods do not match the spatial statistics of measured aero-optic phase screens \cite{UtleyBoiling, UtleyBoiling2}. Although Utley \textit{et al.} \cite{UtleyBoiling2} introduce an anisotropic boiling flow algorithm which generates phase screens that more closely match the spatial correlations of aero-optic data, this model fails to match the temporal power spectrum (TPS) of the measured data.

More recently, Vogel \textit{et al.} \cite{Vogel} and Faghihi \textit{et al.} \cite{Faghihi} proposed data-driven algorithms for generating aero-optic phase screens that do not require the Kolmogorov theory of turbulence and can directly estimate parameters from data. Vogel \textit{et al.} \cite{Vogel} propose a single time-lag vectorized AR model that directly fits the spatial statistics of measured aero-optic phase screen data using Principal Component Analysis (PCA).  However, this approach is unable to model long-range temporal correlations. Faghihi \textit{et al.} \cite{Faghihi} applied a linear state-space model to the top PCA coefficients of measured aero-optic phase screens. This algorithm also does not fully capture long-range temporal correlations, and it relies on a non-linear estimation algorithm from \cite{Chen}. Further, it restricts the synthetic phase screens to a lower-dimensional subset of fewer than 100 principal components, which is not consistent with the behavior of physical data. 

In this paper, we introduce ReVAR (Re-whitened Vector AutoRegression), a data-driven algorithm for generating synthetic data that matches the statistics of measured aero-optic phase screens. ReVAR is computationally efficient for fitting parameters to data and for generating synthetic data.  A key component of ReVAR is a linear predictive model that we call Long-Range AutoRegression, which includes the following novel contributions:
\begin{itemize}
    \item A multiple time-lag AR model to generate synthetic aero-optic phase screen data that matches high-frequency temporal correlations of both the measured data and the streamwise slopes of the measured data.
    
    \item Multiple low-pass filters used to augment this AR model in order to match the long-range temporal correlations of the measured data.
\end{itemize}
Our experiments show that ReVAR has improved error metrics and better matches the TPS of two measured turbulent boundary layer (TBL) data sets \cite{Kemnetz} relative to isotropic and anisotropic boiling flow \cite{Srinath, UtleyBoiling2} and to the single time-lag AR model proposed by Vogel \textit{et al.} \cite{Vogel}. Further, ReVAR better matches the structure function of the measured data than boiling flow and is comparable to the single time-lag AR model.

This paper builds on previous work from \cite{Utley, UtleyBoiling2}. An open-source Python package implementing the ReVAR algorithm is publicly available \cite{ReVAR_Code}.

\section{ReVAR: Algorithm Design}
\label{s: ReVAR: Algorithm Design}

ReVAR takes a time series of measured aero-optic phase screens as input and uses two steps, parameter estimation and data synthesis, to generate synthetic phase screen data. The first step estimates parameters $\hat{\theta}$ of a discrete-time multivariate Gaussian random process whose statistics are time-shift invariant. The second step of ReVAR, data synthesis, uses the parameters $\hat{\theta}$ to generate a sample of this random process. The resulting sample has the same spatial and temporal statistics as the measured aero-optic phase screens. We use the term synthetic aero-optic phase screen data (or synthetic data for short) to describe this generated data.

\subsection{Parameter Estimation from Measured Data}\label{s: Parameter Estimation from Measured Data}

\begin{figure}[ht]
    \centering
    \includegraphics[width=0.8\textwidth]{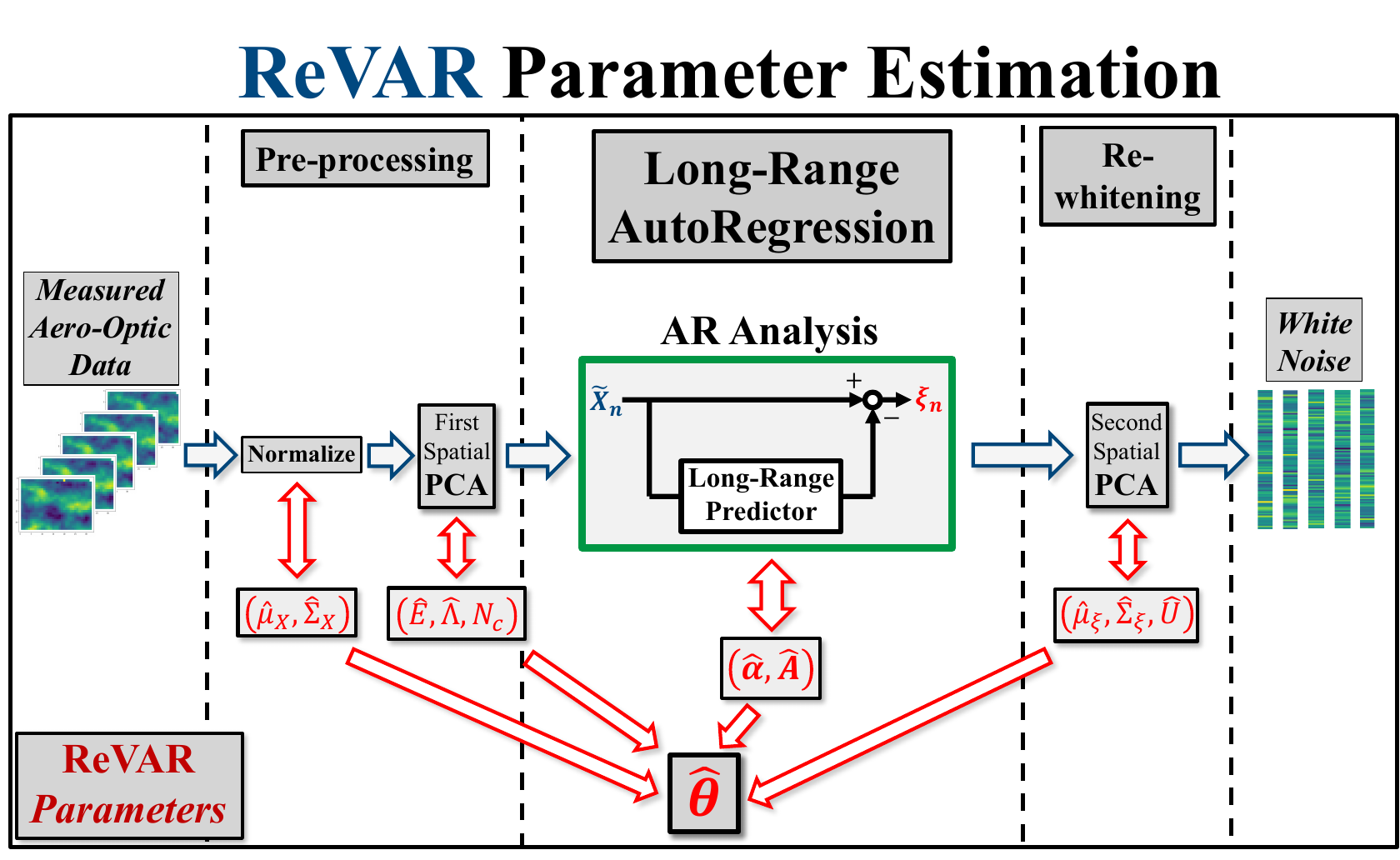}
    \caption{Parameter estimation from measured aero-optic data. This process estimates the parameters $\theta$ of a multivariate Gaussian random process. After normalizing by the time-averaged statistics through Eq.~\eqref{eq: Normalizing} and computing a spatial Principal Component Analysis (PCA) through Eqs.~(\ref{eq: First Spatial PCA}-\ref{eq: Top Principal Coefficients}), ReVAR trains a linear predictive model defined through Eq.~\eqref{eq: Long-Range AR Model}, which we call Long-Range AutoRegression (AR); Long-Range AR uses the long-range predictor of Eq.~\eqref{eq: Long-Range Predictor} (illustrated in Fig.~\ref{fig: Long-Range Predictor}). Finally, ReVAR takes the residuals of the Long-Range AR model and computes a second spatial PCA using Eq.~\eqref{eq: Second Spatial PCA}. Importantly, this converts the residuals to white noise through Eq.~\eqref{eq: Spatial Whitening}, so we can reverse this process with white noise input to generate synthetic aero-optic data.}
    \label{fig: Parameter Estimation}
\end{figure}

Figure~\ref{fig: Parameter Estimation} shows the three steps of parameter estimation from measured aero-optic phase screens: pre-processing, Long-Range AutoRegression, and Re-whitening. Here, we refer to the measured aero-optic phase screens as measured aero-optic data or measured data for short.
\begin{itemize}
    \item {\bf Pre-processing}: Normalize the measured data by removing its sample mean and standard deviation, then compute a spatial Principal Component Analysis (PCA) of the normalized data. This step outputs the principal coefficients of the measured data.
    
    \item {\bf Long-Range AutoRegression}:
    Fit the temporal statistics of the measured data by using the principal coefficients representation to train a linear predictive model, which we call Long-Range AutoRegression. This step outputs the time series of residuals of the Long-Range AR model, which are temporally un-correlated but spatially correlated.
    
    \item {\bf Re-whitening}: Fit the spatial statistics of the measured data by computing a second spatial PCA from the residuals of the Long-Range AR model. The output of this step is spatially and temporally un-correlated white noise.
\end{itemize}
These steps estimate parameters
\begin{align}\label{eq: ReVAR Parameters}
    \hat{\theta} =\Bigl(\underbrace{(\hat{\mu}_X, \hat{\Sigma}_X), (\hat{E}, \hat{\Lambda}, N_c)}_{\text{Pre-Processing}}, \underbrace{(\hat{\bm{\alpha}}, \hat{\bm{A}}),}_{\text{Long-Range AR}} \hspace{0.3cm} \underbrace{(\hat{\mu}_\xi, \hat{\Sigma}_\xi, \hat{U})}_{\text{Re-whitening}}\Bigr)
\end{align}
that determine the spatial and temporal statistics of a Gaussian random process.

By taking white noise input and reversing these steps, we can then generate synthetic data with the same statistics as the measured data.

\subsubsection{Pre-Processing}\label{s: Pre-Processing}
Before fitting the statistics of the measured data, we normalize the spatial distribution of the data and reduce the spatial dimensionality of the data. These steps i) ensure that the parameters of ReVAR do not depend on any units present in the measured data and ii) reduce computational expense in the linear predictive model.

We first reshape the measured data images into vectors. We denote the $n$th frame of the measured data by $X^{\text{meas}}_n\in\mathbb{R}^{N_p}$, where $N_p$ is the number of pixels in each image. Here, the time-index $n$ takes values from $0$ to $N_T-1$, where $N_T$ is the number of time-steps of measured data used for parameter estimation.

\paragraph{Data normalization:} To remove units from the measured data and normalize the data's spatial distribution, we normalize by the pixel-wise sample mean and standard deviation of the data, $\hat{\mu}_X, \hat{\sigma}_X \in \mathbb{R}^{N_p}$. After computing these statistics, we take the diagonal matrix $\hat{\Sigma}_X = \text{diag}(\hat{\sigma}_X)$ and compute the normalized $n$th frame as
\begin{align}\label{eq: Normalizing}
    X_n = \hat{\Sigma}_X^{-1}(X^{\text{meas}}_n - \hat{\mu}_X). \hspace{1cm} \text{(Normalized Data)}
\end{align}

\paragraph{Spatial dimensionality reduction:}
To reduce computational expense in the linear predictive model, we first perform a spatial PCA and restrict to a specified subset of components.
To do this, we use the normalized data to compute the spatial covariance matrix $\hat{R}_X$, where for pixels $i$ and $j$, the $(i,j)$th entry of $\hat{R}_X$ is $(1/N_T)\sum_n X_{n, i} X_{n, j}$.  We then take the Singular Value Decomposition (SVD) of $\hat{R}_X$,
\begin{align}\label{eq: First Spatial PCA}
    \hat{R}_X = \hat{E}\hat{\Lambda} \hat{E}^T, \hspace{1cm} \text{(First Spatial PCA)} 
\end{align}
where $\hat{E}, \hat{\Lambda}\in\mathbb{R}^{N_p\times N_p}$ and the columns of $\hat{E}$ are the orthonormal \textit{principal components}, with variance given by the corresponding diagonal entry $\hat{\lambda}_i$ of $\hat{\Lambda}$. We then represent the normalized data values using this basis of principal components.  This yields the \textit{principal coefficients} \cite{Chatterjee, Berkooz} for the $n$th frame:
\begin{align}\label{eq: Principal Coefficients}
    \tilde{X}_n = \hat{E}^T X_n.\hspace{1cm}\text{(Principal Coefficients)}
\end{align}

We define the prediction subspace as the span of the top $N_c$ principal components that contain 99\% of the normalized data's spatial variance and then project the principal coefficients onto the prediction subspace:
\begin{align}\label{eq: Top Principal Coefficients}
    \tilde{X}_n^{(P)} = P_{N_c}\tilde{X}_n\in\mathbb{R}^{N_c}.\hspace{1cm}\text{(Top $N_c$ Principal Coefficients)}
\end{align}
Here, $P_{N_c}\in \mathbb{R}^{N_c\times N_p}$ extracts the first $N_c$ components of the vector $\tilde{X}_n$.

This choice of prediction subspace allows the linear predictive model to preserve the spatial statistics of the normalized data while reducing computation time in synthetic data generation.

\subsubsection{Long-Range AutoRegression}\label{s: Long-Range AutoRegression}
In this section, we introduce Long-Range AutoRegression, which uses two low-pass filters to capture long-range temporal effects that are not captured by a standard AR model with fixed time lags \cite{Lutkepohl}.

Appendix~\ref{appendix: Long-Range AR Theory} describes the theoretical foundation of the Long-Range AR model and compares Long-Range AR to the standard vectorized AR model and to the AutoRegressive Moving Average (ARMA) model. Importantly, Long-Range AR differs significantly from ARMA in that ARMA applies a moving average to the residual vectors $\xi_n$, whereas Long-Range AR instead applies linear prediction to low-pass filters of the data itself and assumes that the residuals of the model are temporally un-correlated (see Appendix~\ref{appendix: Long-Range AR Theory} for further details).

\begin{figure}[htbp]
    \centering
    \includegraphics[width=0.5\textwidth]{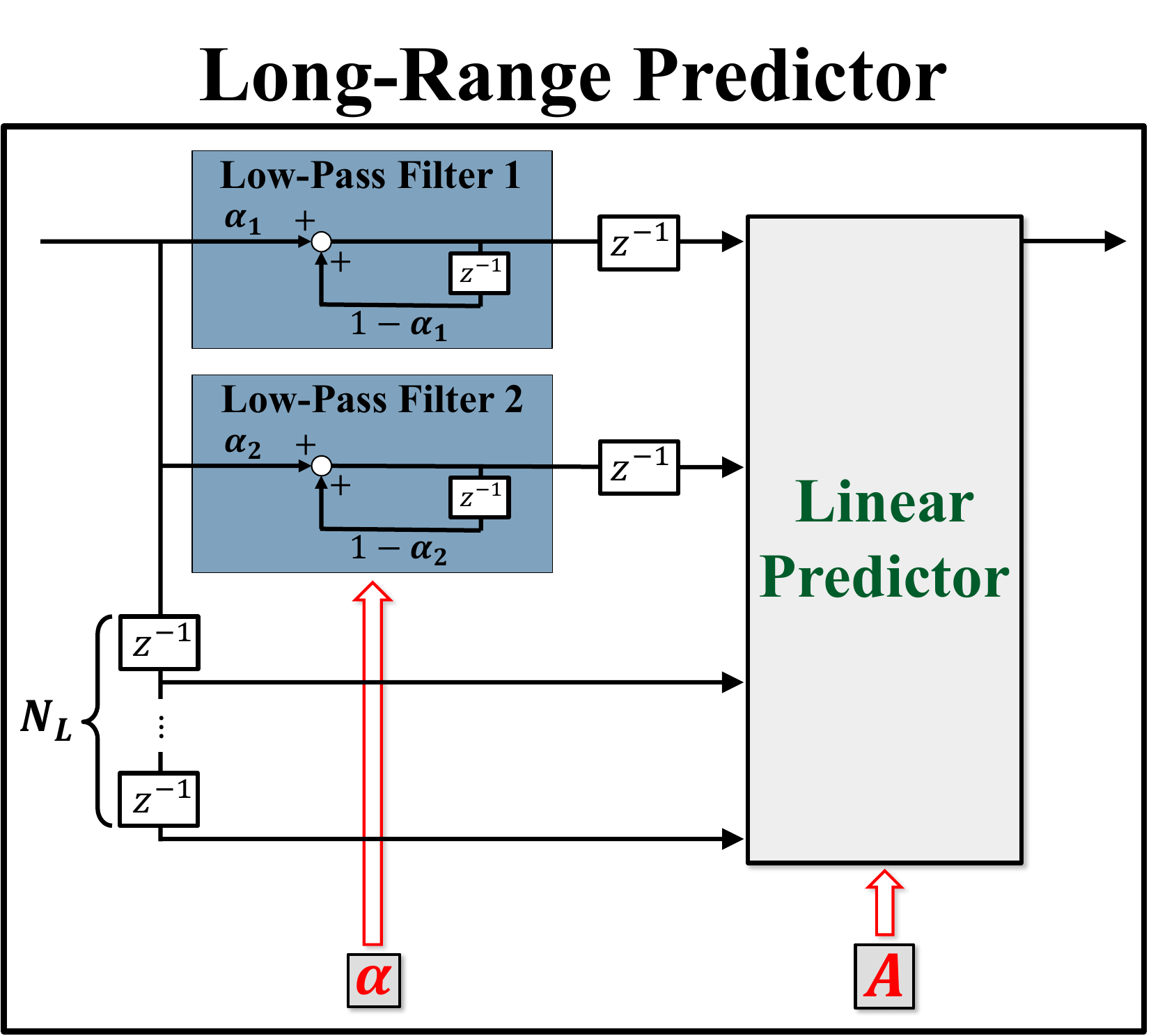}
    \caption{Long-range predictor corresponding to Eq.~\eqref{eq: Long-Range Predictor}. This predictor takes in a time history of data and extracts two components: i) the previous $N_L$ time-steps and ii) two low-pass filters of the entire time series, computed through Eq.~\eqref{eq: Low-Pass Filter}. The predictor then takes a linear combination of these two components with weights $\bm{A}$ to estimate the next vector. We use the low-pass filters to introduce long-range (equivalently, low-frequency) temporal correlations in data generated by this predictor without increasing the number of time-lags $N_L$ to arbitrarily large values. We use two distinct low-pass filters to increase the low-frequency resolution of this predictor.}
    \label{fig: Long-Range Predictor}
\end{figure}

\paragraph{Long-range temporal effects:} To capture long-range (or, equivalently, low-frequency) temporal effects, we compute two causal low-pass filters of the top $N_c$ principal coefficients $\tilde{X}_n^{(P)}$,
\begin{align}\label{eq: Low-Pass Filter}
    Y_{i,n} = (1-\alpha_i)\: Y_{i,n-1}+\alpha_i\:\tilde{X}_n^{(P)},
\end{align}
where $i\in\{1,2\}$ and $Y_{i,n}=\bm{0}$ for $n<0$. Each filter is a first-order low-pass IIR filter (exponential moving average) with transfer function $H(z)={\alpha_i}/({1-(1-\alpha_i)z^{-1}})$, unit DC gain, and a single pole at $z=1-\alpha_i$, yielding an effective cut-off frequency proportional to $\alpha_i$.  We send these low-pass filters as input to a linear predictor, as illustrated in Fig.~\ref{fig: Long-Range Predictor}. The values $\bm{\alpha}=(\alpha_1, \alpha_2)$ determine the cut-off frequencies, and therefore bandwidths, of these low-pass filters. We estimate $\bm{\alpha}$ by specifying two cut-off frequencies and deriving the corresponding values $(\alpha_1, \alpha_2)$, as discussed in greater detail in Appendix~\ref{appendix: Estimating the Low-Pass Filter Parameters}.

We use multiple low-pass filters with different cut-off frequencies to increase the low-frequency resolution of Long-Range AR. Since we represent low-frequency temporal corrections using a basis of low-pass filters of the data, increasing the number of basis elements allows us to increase the frequency resolution of this representation. We found empirically that using two low-pass filters was sufficient for capturing low-frequency temporal effects for the data sets used in this paper, but the method could be applied to any number of filters.

\paragraph{Next-state prediction:}
We combine the low-pass filters from Eq.~\eqref{eq: Low-Pass Filter} with a standard vectorized autoregression to produce a linear prediction of the next state from prior states. The use of low-pass filters introduces long-range temporal correlations in data generated by this linear next-state prediction, so we refer to this predictor as a \textit{long-range predictor}. 

Figure~\ref{fig: Long-Range Predictor} illustrates the long-range predictor. Given a value $N_L\geq 1$, called the \textit{number of time-lags}, this predictor is given by
\begin{align}\label{eq: Long-Range Predictor}
    \hat{X}_n(\bm{A}) = \underbrace{\sum_{\ell=1}^{N_L}A_{X,\ell} \tilde{X}_{n-\ell}^{(P)}}_{\text{AR Component}}\ \ \ \ + \underbrace{A_{Y,1}\hspace{0.05cm} Y_{1,n-1} + A_{Y,2}\hspace{0.05cm} Y_{2,n-1}}_{\text{Low-Pass Filter Component}}. \hspace{1cm}\text{(Long-Range Predictor)}
\end{align}
Here, $\bm{A}$ refers to the set of matrices $A_{X,\ell}$ and $A_{Y,1}, A_{Y,2}$, called the \textit{prediction weights}. The AR component of Eq.~\eqref{eq: Long-Range Predictor} captures and generates short-range temporal correlations, with the specific range restricted by the value of $N_L$. Incorporating a low-pass filter component in this predictor allows us to capture and generate long-range temporal correlations without increasing $N_L$ to arbitrarily large values. The form of Eq.~\eqref{eq: Low-Pass Filter} and the one-step time-delay of the low-pass filters in Eq.~\eqref{eq: Long-Range Predictor} ensure that this long-range predictor is \textit{strictly} causal, so that we can use Eq.~\eqref{eq: Long-Range Predictor} to generate synthetic data.

To generate synthetic data using Eq.~\eqref{eq: Long-Range Predictor}, we first need to estimate the matrices $\bm{A}$ from the measured data. For this, we use a least-squares fit from $N_L$ to the end of the time series ($N_T$):
\begin{align}\label{eq: Prediction Weights}
    \hat{\bm{A}} &= \underset{\bm{A}}{\text{argmin}} \Biggl\{\sum_{n=N_L}^{N_T-1}\Bigl\|\tilde{X}_n^{(P)}-\hat{X}_n(\bm{A})\Bigr\|^2\Biggr\}.
\end{align}
Here, $\hat{X}_n$ is the long-range predictor from Eq.~\eqref{eq: Long-Range Predictor} and $\tilde{X}_n^{(P)}$ are the top $N_c$ principal coefficients, which were computed through Eq.~\eqref{eq: Top Principal Coefficients}. Details on our algorithm for solving Eq.~(\ref{eq: Prediction Weights}), including its computational expense, are discussed in Appendix~\ref{appendix: Long-Range AR Training}.

We note that computation of Eq.~(\ref{eq: Long-Range Predictor}) is of complexity $O\bigl(N_c^2 (N_L+2)\bigr)= O\bigl(N_c^2 N_L\bigr)$ when assuming a fixed number of low-pass filters. Hence, by applying linear prediction to the top $N_c$ principal coefficients $\tilde{X}_n^{(P)}\in\mathbb{R}^{N_c}$ instead of the full principal coefficient vectors $\tilde{X}_n\in\mathbb{R}^{N_p}$, we reduce the computational expense of generating data using Eq.~(\ref{eq: Long-Range Predictor}) by a factor of $N_p^2 / N_c^2$.  In the data presented below, this is more than a factor of four.

\paragraph{Long-Range AR model:}
We use the long-range predictor from Eq.~\eqref{eq: Long-Range Predictor} together with additive noise to generate synthetic data.  Since the long-range predictor is restricted to the top $N_c$ components, we first pad with zeros and then add noise. Hence, we model the principal coefficients $\tilde{X}_n$ from Eq.~\eqref{eq: Principal Coefficients} as a \textit{Long-Range AR process},
\begin{align}\label{eq: Long-Range AR Model}
    \tilde{X}_n &= P_{N_c}^T\hat{X}_n(\bm{A}) + \xi_n, \hspace{1cm} \text{(Long-Range AR)}
\end{align}
where $\hat{X}_n$ is the long-range predictor of Eq.~\eqref{eq: Long-Range Predictor} and $P_{N_c}^T\in\mathbb{R}^{N_p\times N_c}$ pads with zeros. In the generation process described by Eq.~\eqref{eq: Long-Range AR Model}, the 
vectors $\xi_n$ are sampled from a distribution that must be estimated from the measured data.  

During parameter estimation, the $\xi_n$ appear as \textit{residuals} of the Long-Range AR model.  We use the prediction weights $\hat{\bm{A}}$, as estimated through Eq.~\eqref{eq: Prediction Weights}, to compute these residuals as
\begin{align}\label{eq: Residuals}
    \xi_n(\hat{\bm{A}}) = \tilde{X}_n - P_{N_c}^T\hat{X}_n(\hat{\bm{A}}).
\end{align}
Here, we use the notation $\xi_n(\hat{\bm{A}})$ to emphasize the dependence of the residuals on the prediction weights $\hat{\bm{A}}$. 

Given the assumptions of the Long-Range AR model as in  Appendix~\ref{appendix: Long-Range AR Theory} and the computation of $\hat{\bm{A}}$ using a least-squares fit, the residuals $\xi_n(\hat{\bm{A}})$ are temporally un-correlated \cite{Lutkepohl, Bouman}. However, they may be spatially correlated, which we address in the next section.

\subsubsection{Re-whitening}\label{s: Re-Whitening}
Since the residuals $\xi_n(\hat{\bm{A}})$ computed through Eq.~\eqref{eq: Residuals} may be spatially correlated, we use a second spatial PCA to fully de-correlate these residuals. That is, we take the residuals and compute the sample mean vector $\hat{\mu}_\xi$, spatial covariance matrix $\hat{R}_\xi$, and SVD of $\hat{R}_\xi$:
\begin{align}\label{eq: Second Spatial PCA}
    \hat{R}_\xi = \hat{U} \hat{\Sigma}_\xi \hat{U}^T. \hspace{1cm}\text{(Second Spatial PCA)}
\end{align}
Importantly, the time series of vectors
\begin{align}\label{eq: Spatial Whitening}
    W_n(\hat{\bm{A}}) = \hat{\Sigma}_\xi^{-1/2} \hat{U}^T \Bigl(\xi_n(\hat{A}) - \hat{\mu}_\xi\Bigr)
\end{align}
is spatially and temporally un-correlated \textit{white noise} \cite{Berkooz}. For this reason, we call the final step of parameter estimation \textit{Re-whitening}.

\subsubsection{Parameter Estimation Algorithm}\label{s: Parameter Estimation Algorithm}
The process we have outlined in Sec.~\ref{s: Parameter Estimation from Measured Data} converts the measured data $X_n^{\text{meas}}$ to white noise through Eqs.~(\ref{eq: Normalizing},~\ref{eq: Principal Coefficients},~\ref{eq: Long-Range Predictor},~\ref{eq: Residuals}, and~\ref{eq: Spatial Whitening}). In more detail, ReVAR applies the following steps to estimate the parameters $\theta$ from Eq.~\eqref{eq: ReVAR Parameters}:

\begin{enumerate}
    \item Determine the sample mean and standard deviation vectors $(\mu_X, \sigma_X)$ from data and normalize the data through Eq.~\eqref{eq: Normalizing}. This step estimates the normalization parameters $(\hat{\mu}_X, \hat{\Sigma}_X)$ and outputs the normalized data $X_n$.

    \item Take the normalized data $X_n$, compute a spatial PCA as in Eq.~\eqref{eq: First Spatial PCA}, and project the data onto the basis of principal components through Eqs.~(\ref{eq: Principal Coefficients},~\ref{eq: Top Principal Coefficients}). This step estimates the first spatial PCA parameters $(\hat{E}, \hat{\Lambda}, N_c)$ and outputs the principal coefficients $\tilde{X}_n$.

    \item Take the principal coefficients $\tilde{X}_n$ and estimate both the low-pass filter parameters $\bm{\alpha}$ using Appendix~\ref{appendix: Estimating the Low-Pass Filter Parameters} and the prediction weights $\bm{A}$ through Eq.~\eqref{eq: Prediction Weights}. Then, compute the residuals of the Long-Range AR model, $\xi_n(\hat{\bm{A}})$, through Eq.~\eqref{eq: Residuals}. This step estimates the Long-Range AR parameters $(\hat{\alpha},\hat{\bm{A}})$ and outputs the residuals $\xi_n(\hat{\bm{A}})$.

    \item Take the residuals $\xi_n(\hat{\bm{A}})$ and compute the second spatial PCA as in Eq.~\eqref{eq: Second Spatial PCA}. This step estimates the second spatial PCA parameters $(\hat{\mu}_\xi, \hat{\Sigma}_\xi, \hat{U})$ and can be used to convert the residuals to white noise $W_n$ through Eq.~\eqref{eq: Spatial Whitening}.
\end{enumerate}

Since this process converts the measured data to white noise, we can then take white noise input and invert Eqs.~(\ref{eq: Normalizing},~\ref{eq: Principal Coefficients},~\ref{eq: Long-Range Predictor},~\ref{eq: Residuals}, and~\ref{eq: Spatial Whitening}) using the estimated parameters $\hat{\theta}$ to generate synthetic data. 

\subsection{Data Synthesis}\label{s: Data Synthesis}
\begin{figure}[tbp]
    \centering
    \includegraphics[width=0.8\textwidth]{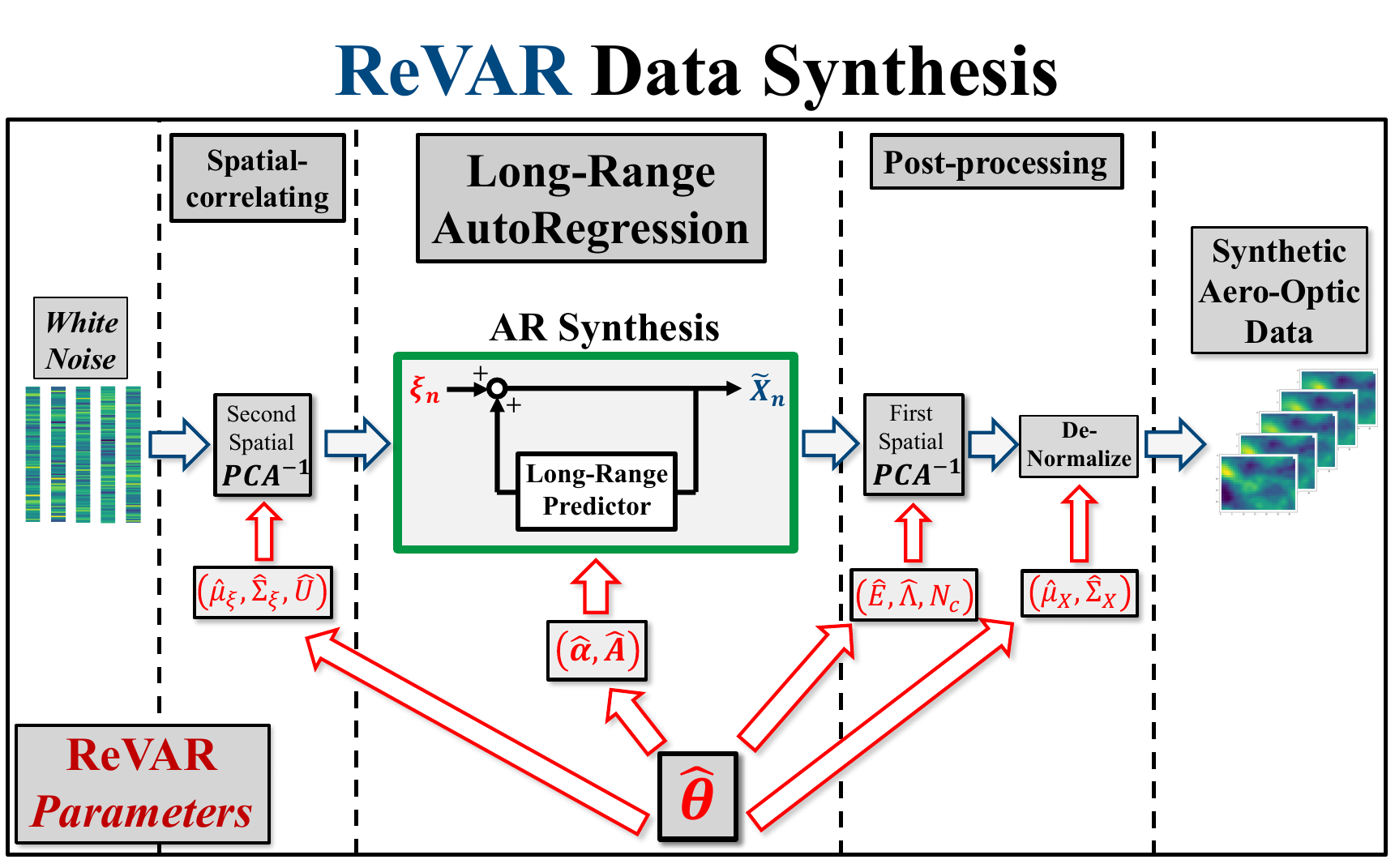}
    \caption{Diagram of ReVAR synthesis. This procedure samples from a multivariate Gaussian random process with the parameters $\hat{\theta}$ from Eq.~\eqref{eq: ReVAR Parameters} by taking white noise input and applying the inverse of the transformation illustrated in Fig.~\ref{fig: Parameter Estimation} through Eqs.~(\ref{eq: Spatially-Correlated Noise},~\ref{eq: Low-Pass Filter},~\ref{eq: Long-Range Predictor},~\ref{eq: Long-Range AR Model}, and~\ref{eq: Generating Synthetic Data}). The synthetic aero-optic phase screen data matches the statistics of the measured data.}
    \label{fig: Synthesis}
\end{figure}

Figure~\ref{fig: Synthesis} illustrates the steps of ReVAR synthesis. This process samples from a Gaussian random process with parameters $\hat{\theta}$ from Eq.~\eqref{eq: ReVAR Parameters} by taking white noise input and applying the inverse of the transformation illustrated in Fig.~\ref{fig: Parameter Estimation}. The inverse transformation applies the following steps:
\begin{itemize}
    \item {\bf Spatial-correlating:} Inverse of Re-whitening. Convert the input white noise to spatially-correlated noise using the second spatial PCA. The output of this step is a time series of vectors that has the same spatial statistics as the residuals of the linear predictive model applied to the measured data from Eq.~\eqref{eq: Residuals}.
    \item {\bf Long-Range AR Synthesis:} Inverse of AR analysis. Convert the spatially-correlated noise to temporally-correlated data by applying the linear predictive model. The output of this step has the same statistics as the principal coefficients of the measured data from Eq.~\eqref{eq: Principal Coefficients}. This step is the inverse of AR analysis from Fig.~\ref{fig: Parameter Estimation}.
    \item {\bf Post-processing:} Inverse of pre-processing. Convert the temporally-correlated data to synthetic data images by inverting the first spatial PCA and denormalizing the data. The output of this process is a synthetic data set with the same spatial and temporal statistics as the measured data.
\end{itemize}

In more detail, we can generate a time series of synthetic data images with any length $N_s$ by inverting Eqs.~(\ref{eq: Normalizing},~\ref{eq: Principal Coefficients},~\ref{eq: Long-Range Predictor},~\ref{eq: Residuals}, and~\ref{eq: Spatial Whitening}). This data generation algorithm applies the following steps:
\begin{enumerate}
    \item Generate $N_s+N_L$ independent white noise vectors $W_n\in\mathbb{R}^{N_p}$ for $-N_L\leq n < N_s$.

    \item For $1 \leq m \leq N_L$, convert the white noise vectors $W_{-m}$ into ``initial'' principal coefficient vectors
    \begin{align}\label{eq: Initial Vectors}
        \tilde{X}_{-m}^{(P)} = P_{N_c}\hat{\Lambda}^{1/2}W_{-m}
    \end{align}
    that have the same spatial statistics as the top $N_c$ principal coefficients computed from measured data through Eq.~(\ref{eq: Top Principal Coefficients}).

    \item Invert the second spatial PCA to convert the white noise $W_n$ to spatially-correlated noise $\xi_n$ for $0 \leq n < N_s$ by modifying Eq.~(\ref{eq: Spatial Whitening}) into
    \begin{align}\label{eq: Spatially-Correlated Noise}
        \xi_n = \hat{U} \hat{\Sigma}_\xi^{1/2} W_n + \hat{\mu}_\xi.
    \end{align}    

    \item Recursively apply the long-range predictor to compute $\hat{X}_n(\hat{\bm{A}})$ for $0 \leq n < N_s$ using Eqs.~(\ref{eq: Low-Pass Filter}-\ref{eq: Long-Range Predictor}).
    
    \item Apply Long-Range AR synthesis to compute \textit{synthetic principal coefficients} $\tilde{X}_n$ for $0 \leq n < N_s$ through Eq.~(\ref{eq: Long-Range AR Model}).

    \item Invert the first spatial PCA and data normalization to convert $\tilde{X}_n$ to synthetic data vectors $X_n^{\text{syn}}$ for $0 \leq n < N_s$ by combining Eqs.~(\ref{eq: Principal Coefficients}) and~(\ref{eq: Normalizing}):
    \begin{align}\label{eq: Generating Synthetic Data}
        X^{\text{syn}}_n = \hat{\Sigma}_X \hat{E} \tilde{X}_n + \hat{\mu}_X.\hspace{1cm}\text{(Synthetic Data)}
    \end{align}
    Reshape the synthetic data vectors $X_n$ into single-channel images.
\end{enumerate}
The output of this process is a time series of $N_s$ single-channel images, each with the same size as the measured data images. Note that since $N_s$ is not specified, the synthetic time series can have arbitrary duration.

\section{Data and Metrics}\label{s: Data and Metrics}
In this section, we describe the measured data along with the metrics used to evaluate our method.

\subsection{Measured Data}\label{s: Measured Data}
We evaluate our method using empirical optical path difference (OPD) data \cite{Kemnetz_Data}, measured through a wind tunnel experiment investigating a TBL. OPD is related to phase aberrations $\phi$ via $OPD=(\lambda/2\pi)\phi$, where $\lambda$ is the wavelength of the detected light waves \cite{Vogel}. Details on the experimental setup, data measurement method using a Shack Hartmann Wavefront Sensor (SHWFS), and post-processing can be found in \cite{Kemnetz, KemnetzDissertation} and Section 5.2 of \cite{UtleyBoiling2}.

Table~\ref{tab: Experimental Data Sets} lists details of the OPD images resulting from this experiment. Here, the OPD data has units of microns, and $N_c$ is determined by retaining 99\% of the variance from the first spatial PCA.

\begin{table}[htbp]
    \caption{Measured Data Sets F06 and F12}
    \label{tab: Experimental Data Sets}
    \centering
    \setlength{\tabcolsep}{2em} 
    \begin{tabular}{l c c}
        \toprule
        \multirow{2}{*}{\textbf{Property}} & \multicolumn{2}{c}{\textbf{Data Set}} \\
        \cmidrule(lr){2-3}
         & \textbf{F06} & \textbf{F12} \\[0.5ex] 
        \midrule
        Number of Pixels, $N_p$ & 735 & 380 \\
        Number of Components, $N_c$ & 264 & 182\\
        Number of Time-Steps, $N_T$ & 150,600 & 251,100 \\
        Sampling Frequency, $f_s$ [kHz] & 100 & 130 \\
        \bottomrule 
    \end{tabular}
\end{table}

\subsection{Quality Metrics} \label{s: Quality Metrics}
We evaluate each method using the temporal power spectrum (TPS) applied to both the OPD and the streamwise slopes of the OPD and using metrics applied to the root mean-square (RMS) amplitude and 2D structure function of the OPD.

The TPS of aero-optic data reveals the temporal frequencies at which the aberrations are most concentrated. This information is essential for training predictive control algorithms, which aid AO correction in specific frequency ranges \cite{Poyneer, Gavel, Paschall}. In fact, some researchers have used the TPS of aero-optic phase aberrations to evaluate the accuracy of predictive control algorithms \cite{Goorskey, Shaffer, ShafferPredictive}. Thus, we compute the TPS of the OPD data (which we call the OPD TPS for short) and denote this quantity by $S_{OPD}$. In addition to the OPD, the streamwise slopes (i.e., the partial derivative of the OPD images in the $x$-direction) are related to the deflection angle $\theta_x$, which is the quantity most directly measured by the SHWFS \cite{Kemnetz, SiegenthalerShear, Cress}. For this reason, we also compute the TPS of these slopes (which we call the slopes TPS) and denote this quantity by $S_{\theta_x}$.
We compute these streamwise slopes using a discrete spatial gradient of the OPD images along the $x$-axis and estimate each TPS using the approach described in Section 5.3 of \cite{UtleyBoiling2}.

In addition to the TPS, the RMS of aero-optic OPD data, often called the $OPD_{rms}$, describes the overall level of aero-optic aberration \cite{WangPhysicsComputation, JumperRecentAdvancements, Visbal} and the structure function $D_{OPD}$ evaluates the spatial statistics of the aero-optic data. We compute $OPD_{rms}$ using an average over both space and time and estimate $D_{OPD}$ using the method described in Section 4 of \cite{UtleyBoiling2}.

We use the normalized root-mean squared error (NRMSE) to compare the TPS and structure function of the measured data with those of the synthetic data. We use the notation NRMSE$(\hat{\bm{y}}, \bm{y}^{\text{data}})$ for the NRMSE between a quantity $\hat{\bm{y}}$ obtained from synthetic data and a quantity $\bm{y}^{\text{data}}$ obtained from measured data. Details on our normalization method and a formula for NRMSE$(\hat{\bm{y}}, \bm{y}^{\text{data}})$ can be found in Section 5.3 of \cite{UtleyBoiling2}.

We report the following errors, where we use the same naming conventions as with $(\hat{\bm{y}}, \bm{y}^{\text{data}})$:
\begin{itemize}
    \item \textbf{Slopes TPS Error}: NRMSE$(\hat{S}_{\theta_x}, S_{\theta_x}^{\text{data}})$. 

    \item \textbf{OPD TPS Error}: NRMSE$(\hat{S}_{OPD}, S_{OPD}^{\text{data}})$.

    \item $\bm{OPD_{rms}}$ \textbf{Error}: The relative error $\big\vert\widehat{OPD}_{rms}-OPD_{rms}^{\text{data}}\big\vert\ /\ OPD_{rms}^{\text{data}}$.

    \item \textbf{Structure Function Error}: NRMSE$\left(\sqrt{\hat{D}_{OPD}}, \sqrt{D_{OPD}^{\text{data}}}\right)$.  
\end{itemize}
Following the approach of \cite{UtleyBoiling2}, we take the square root of $D_{OPD}$. We use this scaling so that both the small and the large structure function values influence the error metric proportionally.

\section{Results}
\label{s: Results}

In this section, we show results from ReVAR applied to measured aero-optic phase screen data. In Sec.~\ref{s: Data Generation Results}, we compare ReVAR with other data-driven approaches. In Sec.~\ref{s: ReVAR Results Across Different N_L Values}, we evaluate ReVAR with various values of $N_L$. We evaluate each method using the error metrics of the previous section. 
In Sec.~\ref{s: ReVAR Computational Expense Results}, we describe the computational advantages of using a prediction subspace. 

\subsection{Parameter Estimation} \label{s: Parameter Estimation}
We estimate the parameters of ReVAR, isotropic and anisotropic boiling flow \cite{Srinath, UtleyBoiling2}, and the single time-lag AR model from Vogel \textit{et al.} \cite{Vogel} from each data set F06 and F12 from Table~\ref{tab: Experimental Data Sets}. For both measured data sets (described in Sec.~\ref{s: Measured Data}), we used the first 80\% of the time series for estimation and used the remaining 20\% for error evaluation. We used the methods from \cite{UtleyBoiling2} to estimate the parameters of isotropic and anisotropic boiling flow, the method in \cite{Vogel} to estimate the parameters of the single time-lag AR model, and the method described in Sec.~\ref{s: Parameter Estimation from Measured Data} with $N_L=4$ to estimate the parameters of ReVAR.

\subsection{Data Generation Results} \label{s: Data Generation Results}
After estimating the parameters for each data set and each method in Section~\ref{s: Parameter Estimation}, we generated corresponding synthetic data. For both data sets, we used each method to generate twenty data sets, each with twenty times the samples of the 20\% error evaluation time series. We computed the errors from Section~\ref{s: Quality Metrics} using these generated data sets versus the 20\% error evaluation data, then averaged over the twenty data sets.

\begin{figure}[ht]
    \centering
    \includegraphics[width=0.8\textwidth]{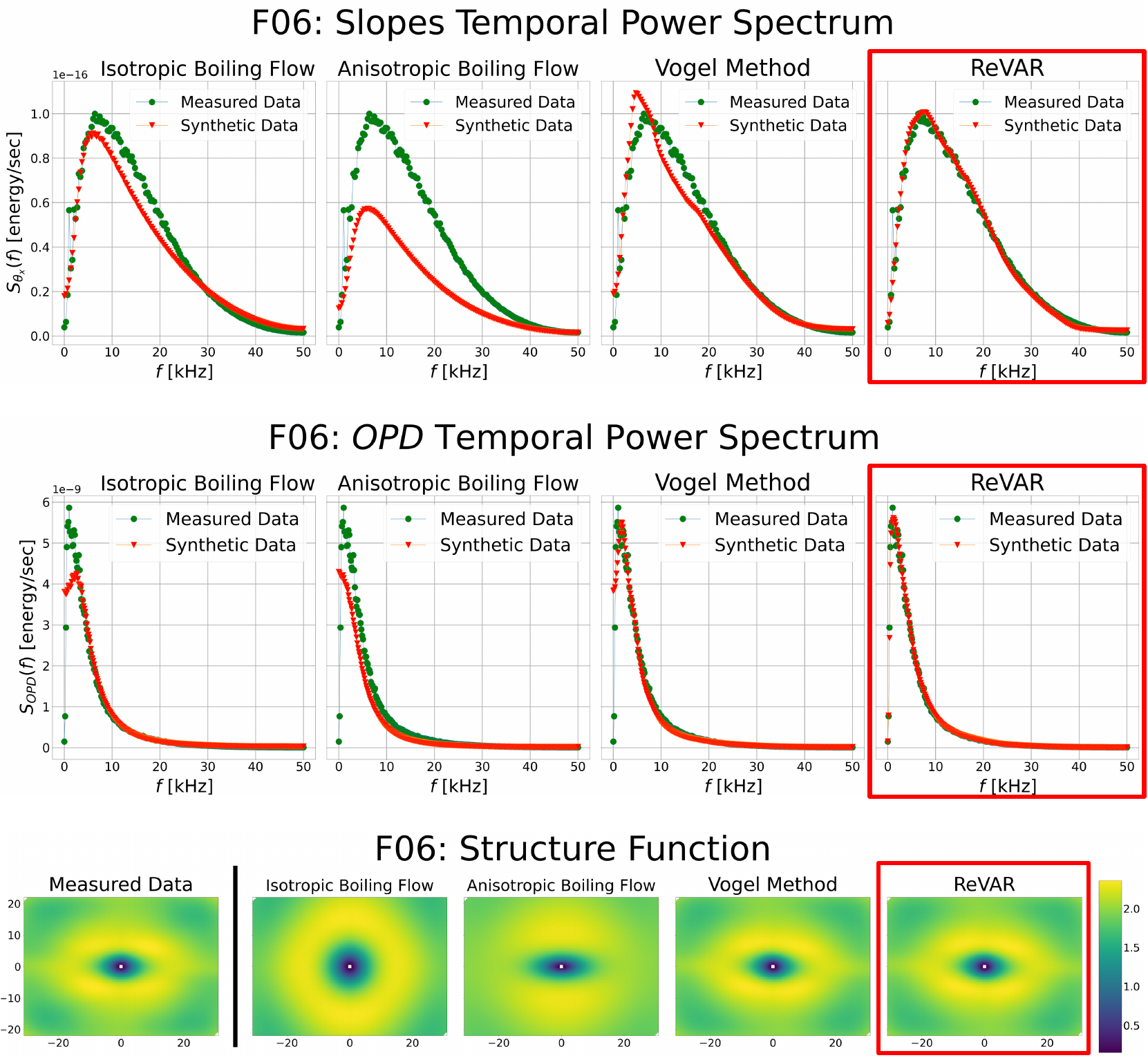}
    \caption{Results from each method for data set F06. {\bf First and second rows:} Comparisons of the slopes and OPD TPS (respectively) obtained from the measured data (green line with circular markers) and synthetic data (orange line with triangular markers), where OPD has units of microns. Boiling flow and Vogel's method both show a mismatch from the measured slopes and OPD TPS at frequencies below 30 kHz and 5 kHz (respectively), while ReVAR (red box) closely matches both measured TPS at all frequencies. {\bf Third Row:} Comparison of the 2D structure functions obtained from measured data (leftmost column) and synthetic data (second-fifth columns). Isotropic boiling flow does not match the anisotropic statistics of the measured structure function and anisotropic boiling flow only roughly matches its contours, while Vogel's method and ReVAR (red box) both closely match the measured structure function.
    }
    \label{fig: F06 Results}
\end{figure}

Figure~\ref{fig: F06 Results} and Table~\ref{tab: F06 Error Metrics} show the TPS and structure function results and list the corresponding error metrics for data set F06. Here, we plot the slopes TPS as a function of frequency $f$ [kHz]; plots of the pre-multiplied slopes TPS as a function of Strouhal number can be found in Appendix~\ref{appendix: TPS Plots with Strouhal Number}. ReVAR closely matches the measured slopes and OPD TPS within 4\% NRMSE, while the remaining methods only match the TPS within 7-26\% NRMSE. Most notably, ReVAR closely matches the measured TPS at low frequencies, whereas the remaining methods do not match the D.C. component of the measured data. Correspondingly, ReVAR matches the $OPD_{rms}$ with 0.13\% relative error, while the remaining methods have $OPD_{rms}$ relative errors between 1-13\%. Lastly, Vogel's method and ReVAR both closely match the measured structure function within 3.2\% NRMSE. In contrast, both boiling flow methods have structure function NRMSE exceeding 25\%. Isotropic boiling flow instead gives an isotropic fit to the measured structure function, while anisotropic boiling flow roughly fits its contours.

\begin{table}[hb]
    \caption{Error Metrics - Data Set F06}
    \label{tab: F06 Error Metrics}
    \centering
    \begin{tabular}{l c c c c}
        \toprule
        \multirow{2}{*}{\textbf{Error}} & \multicolumn{4}{c}{\textbf{Method}} \\
        \cmidrule(lr){2-5}
         & \shortstack{Isotropic\\Boiling Flow} & \shortstack{Anisotropic\\Boiling Flow} & \shortstack{Vogel's\\Method} & {\bf ReVAR} \\[0.5ex] 
        \midrule
        Slopes TPS NRMSE (\%) & \phantom{0}8.35 & 26.86 & \phantom{0}8.27 & \phantom{0}\textbf{3.75} \\
        OPD TPS NRMSE (\%) & \phantom{0}8.79 & 11.62 & \phantom{0}7.44 & \phantom{0}\textbf{1.77} \\
        $OPD_{rms}$ Relative Error (\%) & \phantom{0}1.28 & 13.17 & \phantom{0}1.36 & \phantom{0}\textbf{0.13} \\
        Structure Function NRMSE (\%) & 32.60 & 25.25 & \phantom{0}\textbf{3.18} & \phantom{0}\textbf{3.18} \\
        \bottomrule
    \end{tabular}
\end{table}

\begin{figure}[ht]
    \centering
    \includegraphics[width=0.8\textwidth]{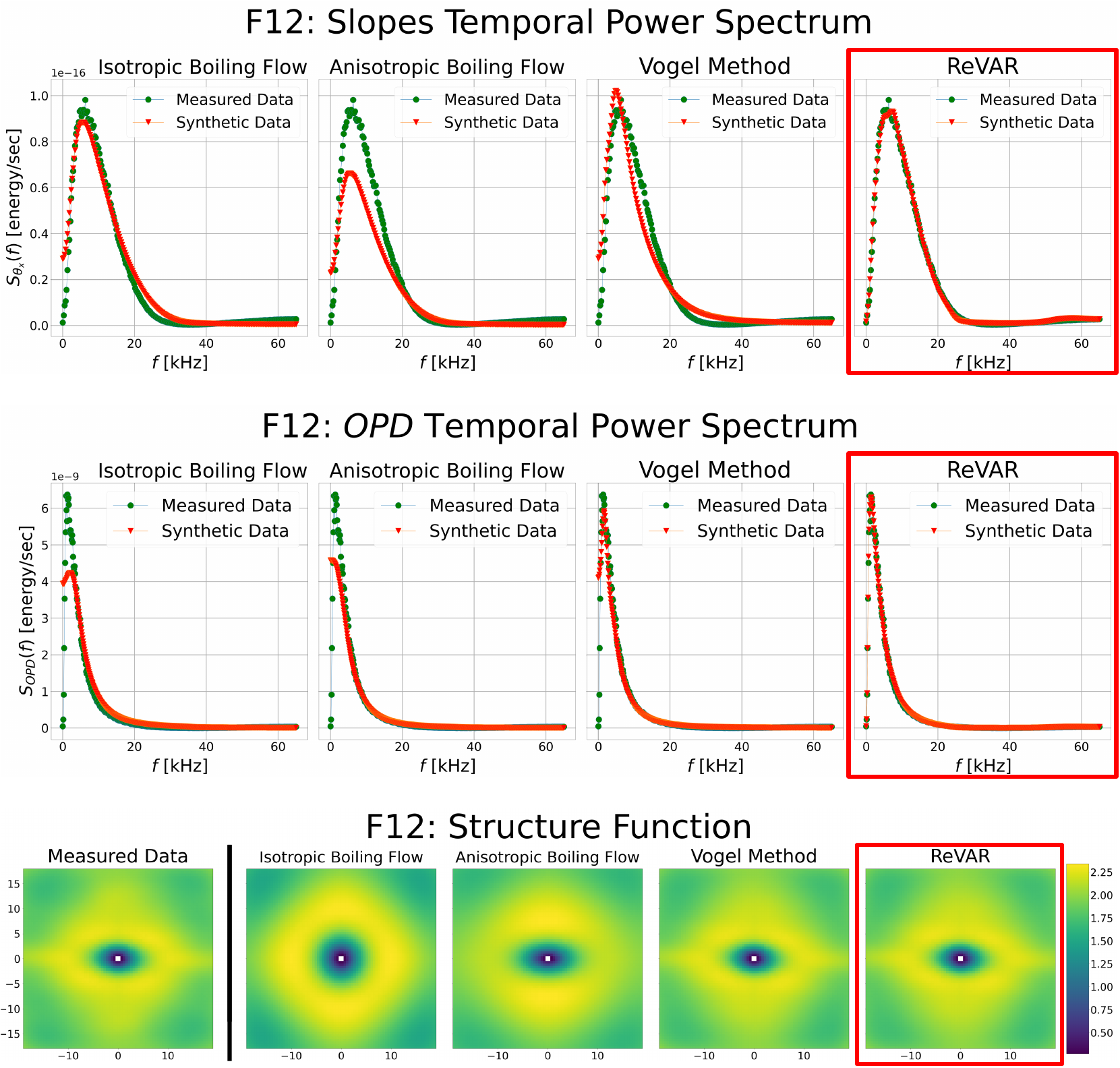}
    \caption{Results analogous to Fig.~\ref{fig: F06 Results} but for data set F12.}
    \label{fig: F12 Results}
\end{figure}

Figure~\ref{fig: F12 Results} and Table~\ref{tab: F12 Error Metrics} show analogous results for data set F12. As with Fig.~\ref{fig: F06 Results}, we plot the slopes TPS as a function of frequency $f$ [kHz]; the pre-multiplied slopes TPS as a function of Strouhal number is plotted in Appendix~\ref{appendix: TPS Plots with Strouhal Number}. Similarly to the results for F06, ReVAR fits both TPS more closely than the remaining methods, especially at low frequencies, and fits the structure function comparably to Vogel's method. Here, ReVAR has TPS NRMSE below 2\% and $OPD_{rms}$ relative error below 0.1\%, while the remaining methods have TPS NRMSE between 5-12\% and $OPD_{rms}$ relative errors between 0.8-4\%. Further, while ReVAR and Vogel's method have structure function NRMSE below 3\%, both isotropic and anisotropic boiling flow's structure function NRMSE exceed 18\%. However, we note that Vogel's method has a slightly lower structure function NRMSE than ReVAR, even after averaging over twenty synthetic data sets. This suggests that Vogel's vectorized AR model implementation, which predicts every vector component \cite{Vogel}, may model spatial statistics marginally better than ReVAR's implementation using a prediction subspace.

\begin{table}[hb]
    \caption{Error Metrics - Data Set F12}
    \label{tab: F12 Error Metrics}
    \centering
    \begin{tabular}{l c c c c}
        \toprule
        \multirow{2}{*}{\textbf{Error}} & \multicolumn{4}{c}{\textbf{Method}} \\
        \cmidrule(lr){2-5}
         & \shortstack{Isotropic\\Boiling Flow} & \shortstack{Anisotropic\\Boiling Flow} & \shortstack{Vogel's\\Method} & {\bf ReVAR} \\[0.5ex]
        \midrule
        Slopes TPS NRMSE (\%) & \phantom{0}5.39 & 11.88 & \phantom{0}8.74 & \phantom{0}\textbf{1.60} \\
        OPD TPS NRMSE (\%) & 11.64 & 12.56 & \phantom{0}9.50 & \phantom{0}\textbf{1.70} \\
        $OPD_{rms}$ Relative Error (\%) & \phantom{0}0.85 & \phantom{0}4.05 & \phantom{0}1.00 & \phantom{0}\textbf{0.05} \\
        Structure Function NRMSE (\%) & 28.65 & 18.17 & \phantom{0}\textbf{2.73} & \phantom{0}2.90 \\
        \bottomrule
    \end{tabular}
\end{table}

\subsection{ReVAR Results as a Function of Number of Lags, $N_L$} \label{s: ReVAR Results Across Different N_L Values}
In this section, we compare TPS NRMSE results from ReVAR with different values of $N_L$, where $N_L$ is the number of time-lags used by the AR component of the long-range predictor in Eq.~\eqref{eq: Long-Range Predictor}. These results motivate our use of $N_L=4$ in Sec.~\ref{s: Data Generation Results}.

Specifically, we evaluate ReVAR with various $N_L$ values from 1 to 6 on both data sets F06 and F12 from Table~\ref{tab: Experimental Data Sets} using the training and data generation methods described in Secs.~\ref{s: Parameter Estimation} and~\ref{s: Data Generation Results}. Here, we use only the slopes and OPD TPS NRMSE described in Sec.~\ref{s: Quality Metrics}. 

\begin{figure}[ht]
    \centering
    \includegraphics[width=0.6\textwidth]{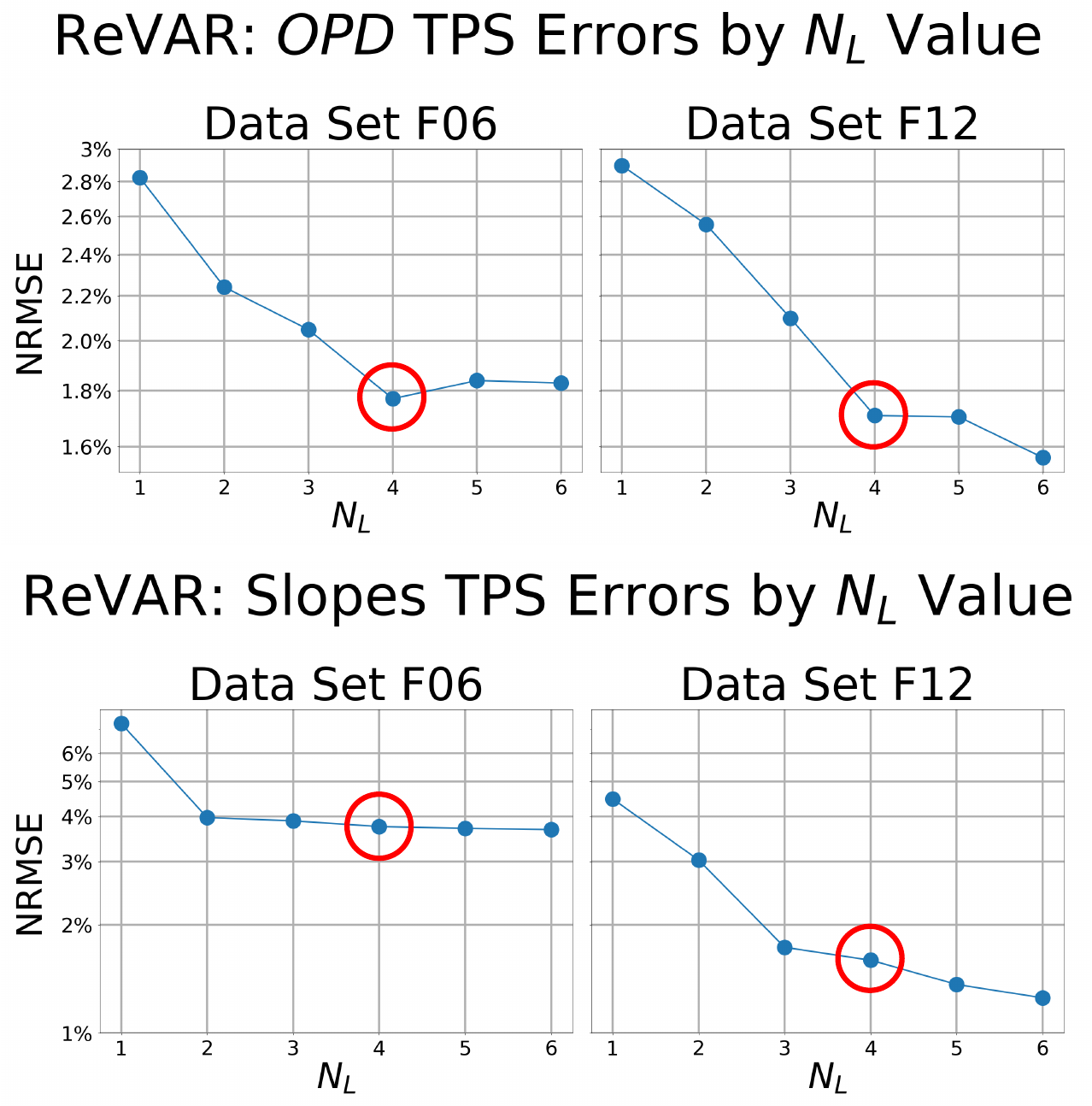}
    \caption{Accuracy of ReVAR as a function of number of time lags, $N_L$.  Each plot shows the TPS NRMSE for various numbers of time-lags $N_L$ for data set F06 (left) and data set F12 (right). {\bf Top:} OPD TPS NRMSE. {\bf Bottom:} Slopes TPS NRMSE. For both data sets, the NRMSE values decrease sharply from $N_L=1$ to $N_L=4$, then either decrease marginally or increase as $N_L$ increases from 4 to 6. To balance this accuracy against the $O(N_L^3)$ computational expense from \eqref{eq:cost of A hat}, we use $N_L=4$ for both data sets.}
    \label{fig: ReVAR NRMSE}
\end{figure}

Figure~\ref{fig: ReVAR NRMSE} shows plots of the slopes and OPD TPS NRMSE across different values of $N_L$. These results show that both TPS NRMSE sharply decrease as $N_L$ increases from 1 to 4, indicating a significant improvement in the TPS fit. However, there is only a marginal improvement in the TPS fit as $N_L$ increases from 4 to 6. For data set F06, the OPD TPS NRMSE increases slightly and the slopes TPS NRMSE remains relatively constant as $N_L$ increases from 4 to 6. Further, for data set F12, the OPD and slopes TPS NRMSE decrease by less than 0.4\% and 1\% NRMSE, respectively. Given the $O(N_L^3)$ computational expense of training a Long-Range AR model from Eq.~\eqref{eq:cost of A hat} and the marginal increase in accuracy when increasing $N_L$ beyond 4, we use $N_L=4$ for both measured data sets.

\subsection{ReVAR Generation Computational Expense Results}\label{s: ReVAR Computational Expense Results}
In this section, we discuss the computational advantage of using a reduced subspace for data generation.  

Recall that we define $N_c$ as the number of principal components required to capture 99\% of the normalized data's spatial variance. For data set F06, the estimated subspace dimension was $N_c=264$, compared to a full pixel count of $N_p=735$. For data set F12, $N_c=182$, compared to $N_p=380$. Thus, the top $N_c$ principal coefficient vectors $\tilde{X}_n^{(P)}$~\eqref{eq: Top Principal Coefficients} have less than half the number of components of the original measured data vectors $X_n^{\text{meas}}$.

Projecting the measured data onto this prediction subspace significantly reduces the computational expense of data generation using a Long-Range AR model. Specifically, the long-range predictor of Eq.~\eqref{eq: Long-Range Predictor} operates entirely within the prediction subspace, applying linear prediction to only the top $N_c$ principal coefficients $\tilde{X}_n^{(P)}$. During data generation, evaluating Eq.~\eqref{eq: Long-Range Predictor} requires several matrix-vector products that scale quadratically with the vector dimensionality. Applying this predictor to vectors of length $N_c$ rather than $N_p$ reduces the computational expense of this synthesis step by a factor of $N_p^2/N_c^2$, which is more than four for both data sets.

This significant reduction in the computational expense of Long-Range AR synthesis heavily outweighs the additional calculations necessary for using the prediction subspace. Namely, our ability to use this prediction subspace relies on the first spatial PCA~\eqref{eq: First Spatial PCA}, which contributes an additional matrix-vector product $\hat{E}\tilde{X}_n$ in Eq.~\eqref{eq: Generating Synthetic Data}. However, since we use $N_L=4$ time-lags and two low-pass filters in Eq.~\eqref{eq: Long-Range Predictor}, the long-range predictor dominates the computational expense of data generation through several matrix-vector products. Thus, reducing the computational expense of Eq.~\eqref{eq: Long-Range Predictor} by more than a factor of four absorbs the extra matrix-vector product in Eq.~\eqref{eq: Generating Synthetic Data}. Accounting for this extra factor, using this prediction subspace reduces the total computational expense of synthetic data generation by more than a factor of two.

\section{Conclusions}
\label{s: Conclusions}

In this paper, we introduced a data-driven algorithm, ReVAR (Re-whitened Vector AutoRegression), for generating synthetic aero-optic phase screen data with statistics that closely match those of measured aero-optic phase screens. ReVAR generates this data computationally and does not encounter the prohibitive monetary or computational expense associated with wind tunnel experiments, flight tests, or computational fluid dynamics simulations. Further, it results in more accurate temporal statistics than existing data-driven approaches by leveraging a Long-Range AutoRegressive (AR) model that fits both short-range and long-range temporal correlations within measured data.

We tested ReVAR on two measured turbulent boundary layer data sets F06 and F12 and compared the results with a single time-lag AR model suggested by Vogel \textit{et al.} \cite{Vogel} and two versions of a conventional phase screen generation method, boiling flow \cite{Srinath, UtleyBoiling2}. Our experiments demonstrate the following through Tables~\ref{tab: F06 Error Metrics} and~\ref{tab: F12 Error Metrics} and Figs.~\ref{fig: F06 Results} and~\ref{fig: F12 Results}:
\begin{itemize}
    \item \textbf{Fit of temporal correlations:} ReVAR closely matches the temporal power spectra (TPS) of both the measured data and its streamwise slopes. Across both data sets, ReVAR matched both TPS within 4\% NRMSE, while the TPS NRMSE of existing methods ranged from roughly 5\% to 27\%. Further, ReVAR captures the TPS at low-frequencies much more closely than these existing methods (see Figs.~\ref{fig: F06 Results} and~\ref{fig: F12 Results}).
    
    \item \textbf{Accuracy of overall variance:} ReVAR matches true $OPD_{rms}$ of both measured data sets within 0.2\% relative error. In contrast, the $OPD_{rms}$ relative errors for existing methods ranged from from 0.8\% to over 13.2\%.
    
    \item \textbf{Fit of spatial correlations:} For both data sets, ReVAR matches the structure function within roughly 3\% NRMSE. These results are comparable to Vogel's method and superior to both boiling flow models (which yielded NRMSE between 18\% and 33\%).
\end{itemize}

Further, the results in Sec.~\ref{s: ReVAR Computational Expense Results} indicate that ReVAR's use of a spatial prediction subspace reduces the computational expense of synthetic data generation by a factor of two for both measured data sets F06 and F12. Thus, ReVAR closely matches the statistics of the measured data sets considered in this paper while maintaining strict computational efficiency. 

\appendix

\section{Long-Range AR: Theoretical Foundation}\label{appendix: Long-Range AR Theory}
In this section, we outline and distinguish standard AR models, ARMA models, and the Long-Range AR model, particularly in their ability to capture long-range temporal correlations. For this, we consider a generic discrete-time multivariate Gaussian random process $X_n$ in a general, purely autoregressive form
\begin{equation}\label{eq: General AR Formulation}
    X_n = \hat{X}_n + \xi_n.\hspace{1cm}\text{(General AR Formulation)}.
\end{equation}
Here, $\hat{X}_n$ is the linear prediction of $X_n$ based on previous data values and $\xi_n$ is temporally un-correlated noise, which we think of as the residuals of the prediction $\hat{X}_n$. 

\paragraph{Standard AR:}
A standard AR model takes Eq.~\eqref{eq: General AR Formulation} and assumes that $\hat{X}_n$ depends linearly on $N_L$ previous data values; the fixed value of $N_L$ is called the number of time-lags \cite{Bouman, Lutkepohl, Pourahmadi}. This results in the standard formulation for the linear prediction in an AR process,
\begin{equation}\label{eq: Standard AR Predictor}
    \hat{X}_n = \sum_{\ell=1}^{N_L}A_\ell X_{n-\ell},\hspace{1cm}\text{(Standard AR Predictor)}
\end{equation}
where $A_\ell$ are the matrices containing the weights of this linear prediction. This predictor uses a finite number of prediction weight matrices $A_1,\dots,A_{N_L}$, all independent of the time-index $n$; these can be estimated efficiently using a least-squares fit to data. 

However, the use of fixed $N_L$ restricts the temporal correlation range captured by these models. One could increase $N_L$ to capture a specified temporal range, but this increases both the computational expense (see Appendix~\ref{appendix: Long-Range AR Training}) and number of prediction weights, which then increases the variance of the estimators of both the prediction weights $A_\ell$ and the linear prediction of $X_n$ \cite{Lutkepohl}.

\paragraph{ARMA:}
An ARMA model uses the standard AR predictor of Eq. \eqref{eq: Standard AR Predictor} for $\hat{X}_n$.  However, it relaxes the assumption that the residuals $\xi_n$ of this prediction are temporally un-correlated and instead introduces a moving average of the past residual vectors $\xi_n$ \cite{Pourahmadi, Lutkepohl}. 
This results in the formulation
\begin{equation}\label{eq: ARMA}
    X_n = \underbrace{\sum_{\ell=1}^{N_L}A_\ell X_{n-\ell}}_{\text{AR Component}} + \underbrace{\sum_{i=1}^{N_i}B_i \ \xi_{n-i}}_{\text{Residual MA}} + \ \xi_n,\hspace{1cm}\text{(ARMA Formulation)}
\end{equation}
where $N_L$ is the number of time-lags as in the standard AR model of~\eqref{eq: Standard AR Predictor} and $N_i$ is a finite number of moving averages.

The moving average component introduces a temporal long-range dependence into the generation of $X_n$ but does not directly relate the long-term temporal statistics of the generated data to the next state prediction.  

\paragraph{Long-Range AR:}
To couple the distant past to the next state prediction more directly, Long-Range AR incorporates a moving average on the states rather than on the residuals. Each average serves as a low-pass temporal filter that can be tuned to approximate a specified temporal frequency cutoff.
In practice, we use two filters but formulate more generally using $N_m$, where each has the form
\begin{align} \label{eq: Moving Average}
    Y_{i, n-1} &= (1 - \alpha_i) Y_{i, n-2} + \alpha_i X_{n-1}. \hspace{1cm}\text{(Low-Pass Temporal Filter)}
\end{align}
The parameter $\alpha_i$ determines the temporal frequency response of each filter and is estimated as in Appendix~\ref{appendix: Estimating the Low-Pass Filter Parameters}. We include these averages with the previous $N_L$ time lags to obtain the next state prediction
\begin{equation}\label{eq: Long-Range AR Formulation}
    X_n = \underbrace{\sum_{\ell=1}^{N_L}A_\ell X_{n-\ell}}_{\text{AR Component}} + \underbrace{\sum_{i=1}^{N_m}B_i \ Y_{i, n-1}}_{\text{Low-Pass Filter}} + \ \xi_n.\hspace{1cm}\text{(Long-Range AR Formulation)}
\end{equation}

Expanding $Y_{i, n-1}$ in Eq. \eqref{eq: Long-Range AR Formulation} shows that $X_n$ depends on the full range $X_0, \ldots, X_{n-1}$ rather than on only the previous $N_L$ states as in Standard AR and ARMA.  
This structure allows Long-Range AR to model long-range temporal correlations without significantly increasing the computation time of training calculations. With $N_m = 2$, only two additional prediction weight matrices must be added to the least-squares fit. This provides a time and memory efficient use of the entire time history to predict the next state of the process.

\section{Estimating the Low-Pass Filter Parameters $\bm{\alpha}$}\label{appendix: Estimating the Low-Pass Filter Parameters}
To estimate $\bm{\alpha}$ in Eq.~\eqref{eq: Low-Pass Filter}, we specify the cut-off frequencies of the associated low-pass filters $Y_{i, n}$. We set the cut-off frequencies to one and two orders of magnitude below the peak of the TPS of the top $N_c$ principal coefficients $\tilde{X}_n^{(P)}$ and derive the weights $\hat{\bm{\alpha}}$ that yield these cut-off frequencies.

More precisely, we first find the peak frequency $\hat{f}_c$ of the TPS of the top $N_c$ principal coefficients $\tilde{X}_n^{(P)}$. For this we use the analytical structure of the amplitude spectrum for a TBL, which increases from a value of zero at D.C. to a unique global maximum \cite{GordeyevSubsonic}. We estimate the TPS of each vector component (in units of energy per time-step) using the approach of \cite{PoyneerExperimental, Oppenheim} by applying a Hamming window on multiple subsets of the time samples and then averaging the resulting TPS estimates over all components and subsets. Importantly, this procedure does not include the (temporal) sampling frequency of the measured data, so the resulting peak location $\hat{f}_c$ has units of cycles per time-step. 

Using techniques similar to \cite{Jones}, we find that by taking $\hat{\alpha}_i = 1-\exp(-2\pi f_i)$, the amplitude of the transfer function for $Y_i$ falls to $1/\sqrt{2}$ at the frequency $\omega=2\pi f_i$ (from a value of 1 at $\omega=0$), thus satisfying the commonly used definition of a cut-off frequency for a low pass filter.

Hence we make a linear approximation to the exponential to choose the values of $\bm{\alpha}$ that result in approximate cut-off frequencies $\hat{f}_c/10$ and $\hat{f}_c/100$:
\begin{align}\label{eq: alpha_0}
    \hat{\alpha}_1 &= \frac{2\pi \hat{f}_c}{10} \approx 1-\exp\Bigl(-\frac{2\pi \hat{f}_c}{10}\Bigr), \\ \label{eq: alpha_1}
    \hat{\alpha}_2 &= \frac{\hat{\alpha}_1}{10}.
\end{align}

\section{Long-Range Predictor: Estimation of Prediction Weights}\label{appendix: Long-Range AR Training}
Recall that the long-range predictor of Eq.~\eqref{eq: Long-Range Predictor} has the form
\begin{align}
    \hat{X}_n(\bm{A}) = \underbrace{\sum_{\ell=1}^{N_L}A_{X,\ell} \tilde{X}_{n-\ell}^{(P)}}_{\text{AR Component}}\ \ \ \ + \underbrace{A_{Y,1}\hspace{0.05cm} Y_{1,n-1} + A_{Y,2}\hspace{0.05cm} Y_{2,n-1}}_{\text{Low-Pass Filter Component}}.
\end{align}
In this section, we describe the least-squares calculations used to estimate the prediction weights $\bm{A}$ and report on the computational expense of these calculations.

\paragraph{Least-squares calculations:} To compute $\hat{\bm{A}}$ through Eq.~(\ref{eq: Prediction Weights}), we estimate the rows of the matrices $A_{X,\ell}$ and $A_{Y,i}$ independently of one another, where $A_{X,\ell}$ and $A_{Y,i}$ are the prediction weight matrices associated with the $\ell$th lag and $i$th low-pass filter, respectively. To do this, we consider a set of $N_c$ least-squares problems, each of size $N_T-N_L$ by $N_c\times (N_L+2)$.

Here, we iterate over the top $N_c$ principal coefficients, indexed by $0 \leq j < N_c$. We denote the $j$th rows of the matrices $A_{X,\ell}$ and $A_{Y,i}$ by $[A_{X,\ell}]_j\in\mathbb{R}^{N_c}$ and $[A_{Y,i}]_j\in\mathbb{R}^{N_c}$, respectively. We can estimate the rows of these matrices independently of one another since the principal coefficient vectors $\tilde{X}_n$ are spatially un-correlated at each time-step \cite{Chatterjee, Berkooz}. This estimation method is derived from the closed-form solution of $\hat{\bm{A}}$ found in \cite{Lutkepohl}.

For each $j$, we define the following objects:
\begin{itemize}
    \item First, we define a vector $[\tilde{X}]_j\in\mathbb{R}^{N_T-N_L}$ containing the time series of the $j$th principal coefficient, as computed from training data:
    \begin{align}
        [\tilde{X}]_j = \begin{bmatrix}
            [\tilde{X}_{N_L}]_j & [\tilde{X}_{N_L+1}]_j & \cdots & [\tilde{X}_{N_T-1}]_j
        \end{bmatrix}^T.
    \end{align}
    Recall that $N_L$ is the number of time-lags and $N_T$ is the number of time-steps of training data. By iterating over the range $N_L \leq n < N_T$, we are taking the full data range (excluding the first $N_L$ time-steps).

    \item Next, define a matrix $Z \in \mathbb{R}^{(N_T-N_L)\times \bigl(N_c(N_L+2)\bigr)}$ containing time series of the top principal coefficients $\tilde{X}_n^{(P)}$ and low-pass filters $Y_{i,n}$:
    \begin{align}
        Z = \begin{bmatrix}
            [\tilde{X}_{N_L-1}^{(P)}]^T  & \cdots & [\tilde{X}_{0}^{(P)}]^T & [Y_{1,N_L-1}]^T & [Y_{2,N_L-1}]^T \\
            [\tilde{X}_{N_L}^{(P)}]^T  & \cdots & [\tilde{X}_{1}^{(P)}]^T & [Y_{1,N_L}]^T & [Y_{2,N_L}]^T \\
            \vdots & \ddots & \vdots & \vdots & \vdots \\
            [\tilde{X}_{N_T-2}^{(P)}]^T  & \cdots & [\tilde{X}_{N_T-N_L-1}^{(P)}]^T & [Y_{1,N_T-2}]^T & [Y_{2,N_T-2}]^T
        \end{bmatrix}.
    \end{align}
    Each row of $Z$ contains previous time-steps of data which are multiplied by the prediction weights to estimate the corresponding value in the vector $[\tilde{X}]_j$. Here, we include ranges of $\tilde{X}_n^{(P)}$ and $Y_{i,n}$ containing $N_T-N_L$ time-steps. Each column corresponding to $\tilde{X}_n^{(P)}$ contains a distinct, shifted temporal window of the data.
    
    \item Last, define a vector $\beta_j\in \mathbb{R}^{N_c(N_L+2)}$ containing the $j$th rows of the (unknown) prediction weight matrices $A_{Y,i}$ and $A_{X,\ell}$:
    \begin{align}\label{eq: Least-Squares Problem - Unknown Quantity}
        \beta_j = \begin{bmatrix}
            [A_{X,1}]_j & \cdots & [A_{X,N_L}]_j & [A_{Y,1}]_j & [A_{Y,2}]_j 
        \end{bmatrix}^T.
    \end{align}
    We solve for these prediction weights $\beta_j$.
\end{itemize}
Since there are many more data points than time lags, we obtain the tall, thin least squares problem
\begin{align}\label{eq: Reduced Optimization Problem}
    \hat{\beta}_j = \underset{\beta_j}{\text{argmin}}\Bigl\{\bigl\|[\tilde{X}]_j - Z\beta_j\bigr\|^2\Bigr\}
\end{align}
and solve this using a conventional least-squares algorithm \cite{LAPACK, VanLoan}. 

Recall that we associated each vector $\beta_j$ to the $j$th rows of the matrices $\hat{A}_{X,\ell}$ and $\hat{A}_{Y,i}$. Thus, after computing the estimates $\hat{\beta}_j$ through Eq.~\eqref{eq: Reduced Optimization Problem}, we use the ordering of the rows $[A_{X,\ell}]_j$ and $[A_{Y,i}]_j$ in Eq.~\eqref{eq: Least-Squares Problem - Unknown Quantity} to place these estimated values into the matrices $\hat{A}_{X,\ell}$ and $\hat{A}_{Y,i}$. This is explicitly given by
\begin{align}
    [\hat{A}_{X,\ell}]_{j,k} &= [\hat{\beta}_j]_{(\ell-1) \times N_c +k}, \\
    [\hat{A}_{Y,i}]_{j,k} &= [\hat{\beta}_j]_{(i-1+N_L)\times N_c+k}.
\end{align}
If we enforce $A_{Y,1} = A_{Y,2} = \bm{0}$ (i.e., we use a standard vectorized AR model with $N_L$ time-lags), this is equivalent to the method of directly computing $\hat{\bm{A}}$ from \cite{Lutkepohl}. 

\paragraph{Computational expense of training:} Assuming that the number of low-pass filters used by the long-range predictor is fixed at two, the values $(N_c, N_L)$ determine the complexity of this computation. Given the complexity of finding each solution $\hat{\beta}_j$ in Eq.~(\ref{eq: Reduced Optimization Problem}), which can be derived from \cite{LAPACK, VanLoan}, the complexity of the full least-squares calculations (for all top principal coefficient indices $0 \leq j < N_c$) is then
\begin{align}
    &\hspace{0.4cm}O\bigl(N_T N_c^3 (N_L+2)^2\: + \:N_c^4 (N_L+2)^3\bigr)\\
    &=O\bigl(N_c^3 N_L^2  (N_T \: + \:N_c N_L)\bigr).
\end{align}
If the number of training time-steps $N_T$ remains fixed, then the complexity of computing $\hat{\bm{A}}$ is
\begin{align} \label{eq:cost of A hat}
O\bigl(N_c^4 N_L^3\bigr).
\end{align}

Applying the long-range predictor of Eq.~\eqref{eq: Long-Range Predictor} to only the top $N_c$ principal coefficients instead of all $N_p$ principal coefficients significantly reduces both the computational expense of the least-squares calculations and the number of parameters of this linear predictive model.

\section{Pre-Multiplied Slopes TPS Plots as a Function of Strouhal Number}\label{appendix: TPS Plots with Strouhal Number}
In this section, we show the slopes TPS results for both measured data sets F06 and F12 in a format more commonly-used in the aero-optics field. 

Specifically, we plot the pre-multiplied slopes TPS, evaluated as a function of the unit-less Strouhal number $\text{St}_{\delta}$. We use the boundary layer thickness $\delta^*$=15.6 mm from \cite{Kemnetz, KemnetzDissertation} and estimate the convective velocities to be $U_c=172.37$ m/s for F06 and $U_c=160.26$ m/s for F12. Further details on the formulation of $\text{St}_{\delta}$ and our methodology for computing $U_c$ can be found in Appendix C of \cite{UtleyBoiling2}.

Figure~\ref{fig: Pre-Multiplied Slopes TPS} shows the resulting power spectra from both data sets and each method. The horizontal axes display the Strouhal number $\text{St}_{\delta}$ on a logarithmic scale and the vertical axes display the pre-multiplied TPS values $\text{St}_{\delta}\times S_{\theta_x}(\text{St}_{\delta})$ on a linear scale. To prevent aliasing artifacts in these plots, we restrict the Strouhal number values in these plots to a maximum of $\text{St}_{\delta}=3$ and $\text{St}_{\delta}=2$ for F06 and F12, respectively. These pre-multiplied TPS plots show a similar fit to the measured slopes TPS as the unscaled results in Figs.~\ref{fig: F06 Results} and~\ref{fig: F12 Results}.

\begin{figure}
    \centering
    \includegraphics[width=0.8\textwidth]{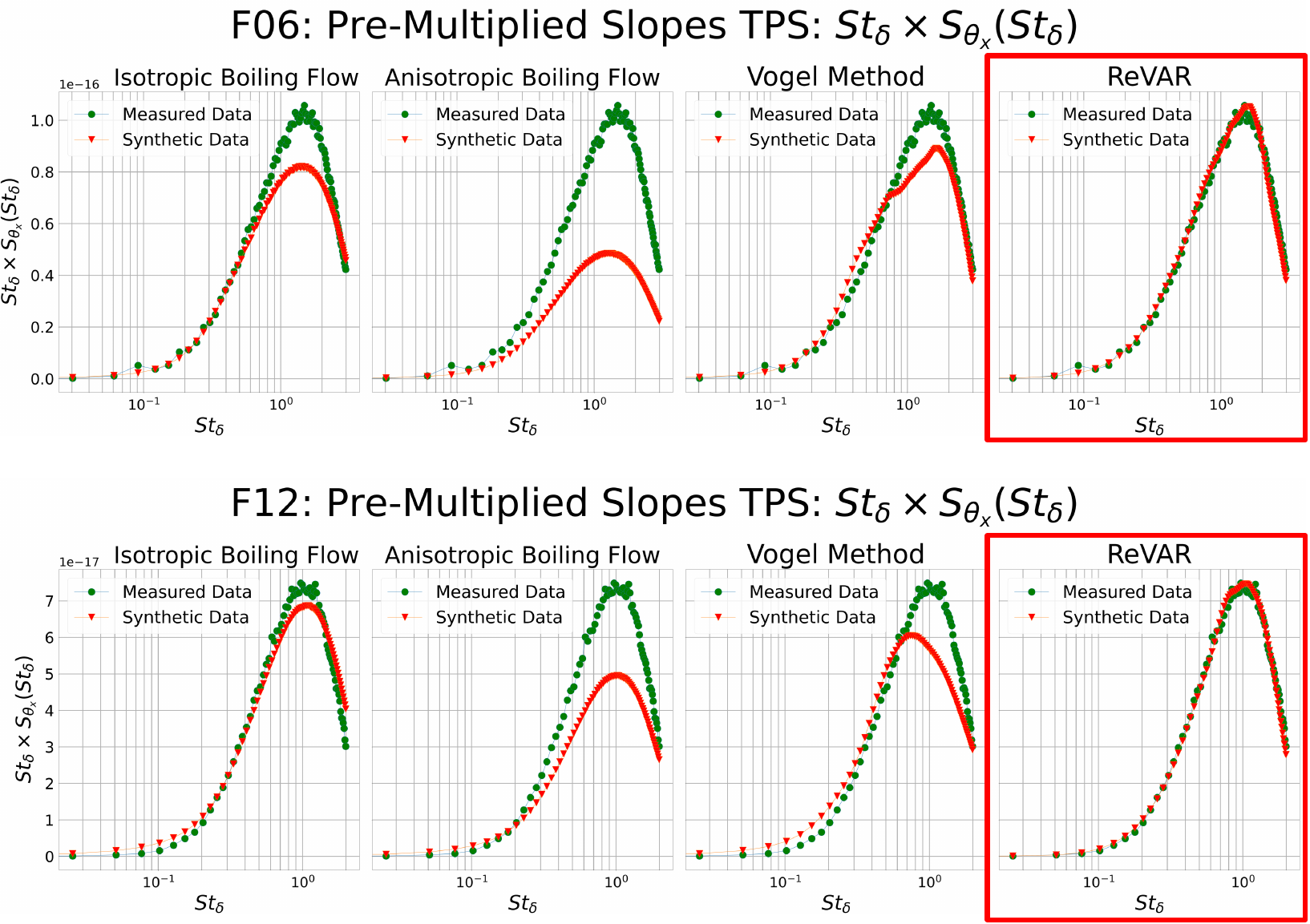}
    \caption{Comparisons of the pre-multiplied slopes temporal power spectrum (TPS) as a function of Strouhal number $\text{St}_{\delta}$ obtained from measured data (green line with circular markers) and synthetic data (orange line with triangular markers) for each method. {\bf Top:} data set F06. {\bf Bottom:} data set F12. These results show consistent accuracy with the standard slopes TPS in Figs.~\ref{fig: F06 Results} and~\ref{fig: F12 Results}.}
    \label{fig: Pre-Multiplied Slopes TPS}
\end{figure}

\begin{backmatter}
\bmsection{Funding}
Showalter Trust, Air Force Research Laboratory (FA9451-20-2-0008).

\bmsection{Acknowledgments}
The authors thank the Showalter Trust and the United States Air Force for supporting this research. The measured data sets used in this article were taken at the University of Notre Dame in the trisonic wind tunnel facility within the Hessert Laboratory for Aerospace Research; for more details concerning the experiment, please refer to Ref.~\cite{Kemnetz}.

\bmsection{Disclosures}
The views expressed are those of the authors and do not necessarily reflect the official policy or position of the Department of the Air Force, the Department of Defense, or the U.S. Government.  Approved for public release; distribution is unlimited.  Public Affairs release approval \# AFRL-2026-0858. The authors declare no conflicts of interest.

\bmsection{Data Availability}
Data underlying the results presented in this paper are publicly available in Ref.~\cite{Kemnetz_Data}. The Python package used to generate the synthetic aero-optic phase screens is publicly available in Ref.~\cite{ReVAR_Code}.

\end{backmatter}

\bibliography{bibliography}

\end{document}